# Attaining the Ground State of Kagome Artificial Spin Ice via Ultrafast Site-Specific Laser Annealing

D. Pecchio,[1,2] S. Sahoo,[1,2,*] V. Scagnoli[1,2,†] and L. J. Heyderman[1,2,‡]

[1]*Laboratory for Mesoscopic Systems, Department of Materials, ETH Zurich, 8093 Zurich, Switzerland*
[2]*PSI Center for Neutron and Muon Sciences, 5232 Villigen PSI, Switzerland*

**ABSTRACT**. Artificial spin ices (ASIs) provide a versatile platform to explore magnetic frustration and emergent phenomena. However, in kagome ASI, experimental access to the ground state remains elusive due to dynamical freezing. Here, we demonstrate a deterministic and rewritable approach to attain the ground state using ultrafast, site-selective laser annealing. By engineering sublattice-dependent optical absorption through selective capping of the nanomagnets with Cr or utilizing different nanomagnet thicknesses, we achieve selective partial demagnetization of one sublattice under a sub-coercive magnetic field, driving the system into the ground state in a single switching step. Magnetic force microscopy reveals nearly perfect long-range ordering, while heat-transfer simulations confirm the sublattice-selective excitation mechanism. This work establishes an ultrafast method to attain the kagome ASI ground state, which does not require a modification of the geometry of the ASI or the materials used for the individual nanomagnets. Beyond ground-state writing, this site-selective activation provides an important tool for controlling the magnetic states, which is important for applications such as reconfigurable magnonic crystals, neuromorphic computing and programmable nanomagnetic logic.

## I. INTRODUCTION

Artificial spin ices (ASIs) are lithographically patterned arrays of dipolar-coupled, single-domain nanomagnets [1], designed to emulate the magnetic frustration and cooperative behavior observed in geometrically-constrained magnetic systems such as those found in the rare-earth pyrochlores [2]. With the possibility to create various geometries lithographically, and therefore tune the interactions, these magnetic metamaterials provide a versatile platform to explore collective phenomena – such as emergent magnetic monopoles [3,4], phase transitions [5–8], and magnetic frustration [9,10] – in a tailored, reconfigurable, and directly observable setting.

Among the various ASIs explored, the kagome ASI stands out due to its high degree of magnetic frustration and the resulting manifold of degenerate low-energy microstates [11–14], and a central objective in the study of kagome ASI is to identify and access its various thermodynamic phases [5,6,12,13]. In this system, stadium-shaped single-domain nanomagnets, whose magnetic moments are referred to as macrospins, are arranged on hexagonal plaquettes so that three nanomagnets meet at each lattice vertex with an angle between them of 120° [see Fig. 1(a) and (b)]. The macrospins associated with each nanomagnet can be denoted as pointing in (pointing out) when their magnetization points towards (away from) the vertex. The so-called "ice rule" refers to satisfying the local magnetostatic constraints at any given vertex, resulting in the lowest energy 1-in/2-out or 2-in/1-out configuration, which is sixfold degenerate, while the doubly degenerate all-in (3-in) or all-out (3-out) vertices form high-energy defects. This can be formalized with the dumbbell model [4,15], where each nanomagnet is replaced by a dumbbell with $\pm q$ magnetic poles at its ends. The vertex magnetic charge is then given by $Q = q \sum_i s_i$ with $s_i = +1$ at the vertex for macrospins pointing in and $s_i = -1$ for macrospins pointing out. Ice-rule-obeying vertices therefore carry the minimum charge magnitude $|Q| = q$, whereas 3-in/3-out defects have $|Q| = 3q$. These $|Q| = 3q$ defects behave as magnetic excitations, which are referred to as emergent magnetic monopoles in an ASI context [3,4,16], although it should be pointed out that ice-rule-obeying vertices can also be associated with such excitations, since these are excitations above the charge ordered ground state, which is described in the next paragraph, and this can be clarified by considering the smeared charge density [4].

* Contact author: sourav.sahoo@psi.ch
† Contact author: valerio.scagnoli@psi.ch
‡ Contact author: laura.heyderman@psi.ch

In terms of the temperature-dependent behavior, theoretical models predict a sequence of magnetic phases on decreasing the temperature [12,13], which are illustrated in Fig. 1(c). Here the arrows on the schematics denote the macrospins, small red (blue) dots correspond to ice-rule-obeying 2-in/1-out (1-in/2-out) vertex configurations, and large red (blue) dots indicate high-energy 3-in (3-out) defects. Going from left to right in Fig. 1(c), at high temperature, the system is in a paramagnetic phase. Then, on cooling, the system first enters a globally disordered but ice-rule-obeying regime known as Spin Ice I. Further cooling promotes the Spin Ice II phase, in which the vertex magnetic charges $Q = \pm q$ arrange in an alternating pattern where every positive (negative) charge is surrounded by three negative (positive) charges, thus minimizing the energy associated with the Coulomb-like interaction between them. This ordered charge configuration emerges despite the disorder in the underlying macrospins, which fluctuate through the reversal of closed, and therefore divergence-free, loops of macrospins, at least theoretically, so that the charge order is maintained. At even lower temperatures, long-range dipolar interactions select a long-range-ordered spin-crystal ground state characterized by alternating clockwise/counterclockwise loops of head-to-tail macrospins on the hexagonal plaquettes surrounding a plaquette with head-to-head/tail-to-tail macrospins.

While this sequence of phases is well-established theoretically, in an experimental system, the magnetization dynamics can become arrested before the ground state is reached. This is a result of dynamical freezing; a collective effect in which the system becomes kinetically trapped in a metastable state before reaching equilibrium [17]. This freezing typically occurs at temperatures above those required to access the Spin Ice II phase. Compounding this issue, the blocking temperature – which marks the onset of thermally activated switching of individual nanomagnets within the experimental timescale – often exceeds the critical temperature associated with collective ordering. As a consequence, the macrospins become magnetically frozen before the system can reach the lower-energy regions of its configuration space.

While indirect characterization methods have provided evidence of low-temperature phase transitions in kagome ASI [5], the corresponding ordered phases have never been observed [18,7]. This has motivated the search for alternative innovative strategies such as exploiting the interfacial Dzyaloshinskii–Moriya interaction (DMI), which reduces the energy barrier to nanomagnet switching. However, this method has only brought a partial charge ordering in kagome ASIs that are disconnected, i.e. made up of arrays of separated nanomagnets [19]. Therefore, up to now, accessing the ground state has typically required breaking the local symmetry of the lattice. In particular, by introducing an asymmetry in the design of the vertices, the degeneracy of vertex configurations can be lifted, and the ground state stabilized. In connected ASIs formed from nanowire networks, this has been achieved through nanoscale notches [17] or asymmetric magnetic bridges [20], enabling direct imaging of the phase transitions that lead to the lowest energy state. In addition, Yue *et al*. [21] demonstrated that ground-state crystallization can be achieved in fully disconnected ASIs by modifying the lengths of specific nanomagnets. However, while effective, this strategy also relies on geometric modifications to the lattice, thus altering the intrinsic frustration.

Alternative methods to achieve the ground state in kagome ASI are based on local manipulation of the magnetization. For example, magnetic force microscopy (MFM) tips have been successfully used to define specific microstates, including the kagome ASI ground state [22]. However, this method is inherently serial, and therefore has a low throughput, making it unsuitable for writing magnetic configurations into large arrays of nanomagnets. Recent advances in exchange bias patterning offer a partial solution to this scalability issue, by spatially modulating the local coercivity in ASIs to allow selective reversal of macrospin subsets with a magnetic field [23]. It is also possible to tailor the width, and thus the shape anisotropy, of selected nanomagnets to deterministically set their switching thresholds. This can be, for example, used to set the chirality of loops of spins in the hexagonal plaquettes of kagome ASI [24]. However, while effective, this approach also relies on introducing an asymmetry in the geometry, and therefore alters the intrinsic frustration of the lattice. These limitations highlight the need for an approach that can address large arrays while preserving the intrinsic frustration, while providing rapid and reliable stabilization of specific states in kagome ASIs, in particular the ground state.

Here, we present a deterministic method for accessing the spin-crystal ground state of kagome artificial spin ice via ultrafast, site-selective laser-induced magnetic relaxation. To this end, we fabricate a kagome ASI composed of disconnected permalloy nanomagnets arranged into two interleaved sublattices [Fig. 1(a)]. While both nanomagnet sublattices are identical in terms of their magnetic properties, they are differentiated by their capping layers: one is capped with Al (in dark gray) while the other is capped with an Al/Cr bilayer (in turquoise). As a result of the optical attenuation by Cr, on illumination of the kagome ASI with an ultrafast laser pulse, the underlying magnetic elements capped with Al/Cr absorb significantly less laser energy than those capped only with Al. This enables precise control of delivery of energy at the sublattice level to attain the ground state without introducing any geometrical asymmetry in the kagome ASI.

By carefully selecting the fluence associated with the ultrafast laser exposure, the nanomagnet sublattice with the Al cap – so without a Cr layer – absorbs sufficient energy to undergo almost complete ultrafast laser-induced demagnetization [25–27]. These thermally activated nanomagnets can then relax and reverse their magnetization under an applied magnetic field, while the elements capped with Al/Cr remain magnetically frozen. The resulting macrospin configuration corresponds to the spin-crystal ground state of the kagome ASI, which we confirm by directly imaging the magnetic configurations with magnetic force microscopy.

We thus demonstrate deterministic and ultrafast access to the kagome ASI ground state over large areas, in a fully disconnected ASI, without relying on structural or material-induced changes to the magnetic properties. This establishes a route to rapidly control the magnetic configurations in ASIs through sublattice-selective laser excitation, without altering the intrinsic magnetic frustration.

## II. SAMPLE PREPARATION AND EXPERIMENTAL SETUP

To experimentally demonstrate this concept, we fabricated kagome ASIs specifically designed to host these two optically distinguishable yet magnetically equivalent sublattices. The kagome ASIs consisted of 100 × 100 μm$^2$ arrays of 8 nm-thick permalloy (Py; Ni$_{83}$Fe$_{17}$) nanomagnets, which included ~84570 nanomagnets. The nanomagnet arrays were patterned using electron beam lithography on silicon (100) substrates in combination with thermal evaporation of permalloy at a base pressure of 6×10$^{-6}$ mbar and lift-off. The nanomagnets had a nominal length and width of $L$ = 300 nm and $W$ = 100 nm, and a lattice constant of $a$ = 320 nm, as illustrated in the scanning electron microscopy image of part of a kagome ASI in Fig. 1(b). Each nanomagnet was initially capped with a 3 nm-thick aluminum layer to prevent oxidation. To create differential optical absorption on the different sublattices while preserving magnetic equivalence, in a second lithography step, an additional 6 nm Cr layer was deposited on one sublattice, resulting in interleaved Py/Al and Py/Al/Cr nanomagnets, as depicted in Fig. 1(a) and further discussed in Section IV.A. This choice of layer stacks for the two sublattices, with the magnetic layer being deposited within the same lithography step, ensured that the only difference between sublattices is in their optical response. The coercive field of the kagome ASI was measured with magneto-optic Kerr effect microscopy to be ~200 Oe.

As an alternative implementation, in a separate set of samples, we also realized kagome ASIs in which both sublattices consisted of Py/Al nanomagnets but with different Py thicknesses (8 nm and 14 nm) defined in two successive lithography steps. This alternative approach is discussed in Section IV.C.

Ultrafast annealing was performed with a femtosecond Nd:YAG pulsed laser (1030 nm fundamental wavelength, 200 kHz repetition rate), frequency-doubled to a wavelength of 515 nm using a barium borate (BBO) crystal and applied at an incident angle of 45°. The laser pulses had a duration of ~100 fs and a focal spot size on the sample of ~85 μm (FWHM). An average power of 45 mW, which corresponded to a fluence of 7.9 mJ/cm$^2$, was applied to give an almost complete demagnetization of the Al-capped nanomagnets, thus allowing their reversal in a sub-coercive field, while the Al/Cr-capped elements remained below the reversal threshold.

The resulting magnetic configurations were imaged ex situ using magnetic force microscopy. The spin maps were reconstructed from the MFM images by analyzing the magnetic contrast, where the two ends of a nanomagnet appear bright and dark, which corresponds to stray field at the north and south pole of each nanomagnet, respectively. For this, using a Python script, the position of each nanomagnet was identified from the corresponding topographic image, and the magnetization direction was assigned by evaluating the gradient in contrast along the nanomagnet long axis.

### III. ULTRAFAST LASER-INDUCED MAGNETIC RELAXATION IN UNIFORM KAGOME ASI

Before introducing sublattice selectivity, we first tested whether ultrafast laser annealing alone could drive the kagome ASI towards low-energy magnetic configurations. For this, we implemented the same decreasing-fluence protocol that we previously developed for square ASIs [27]. This protocol involves exposing the sample to a sequence of femtosecond laser pulses at gradually decreasing fluence, going from 9.7 down to 6.1 mJ/cm$^2$ in steps of 13.2 nJ/cm$^2$, over a 1 s total exposure duration. At high fluence, ultrafast demagnetization and thermal fluctuations are sufficient to activate magnetization reversal in all nanomagnets. We can assume that there is a distribution in the energy barriers to switching of the nanomagnets [4], which is a result of the variability in the nanomagnet shape and microstructure as well as differences in the local magnetic fields. As the laser fluence gradually decreases, it seems likely that nanomagnets with higher energy barriers freeze first, while those with lower energy barriers continue to switch. The frozen magnets can form local nucleation centers that seed the growth of ordered magnetic domains. Dipolar interactions between already-frozen and still-active nanomagnets then guide the relaxation process, promoting the formation of low-energy configurations as the system cools down. In square ASI, this mechanism enables collective relaxation into well-ordered magnetic states [27], achieving a level of order at ultrafast timescales that is comparable or superior to that obtained via conventional thermal annealing [28–30], but without the need for prolonged equilibration protocols. In the current work, the goal is to evaluate whether a such an ultrafast protocol could also be effective in achieving the ground state for the more frustrated geometry of the kagome ASI.

Following laser annealing, an 8 × 8 µm² region of the kagome ASI, including ~540 nanomagnets, was imaged using MFM as shown in Fig. 2(a). The size of this region was chosen to ensure that all of the nanomagnets were exposed to a laser fluence that is sufficiently high to demagnetize them and therefore allow them to switch. The macrospin orientations were reconstructed from the magnetic contrast in the 8 × 8 µm² area, yielding the magnetic configuration [Fig. 2(b)]. From this reconstructed configuration, we determined the charge-charge correlators [18,31], which give an indication of the degree of magnetic ordering. While the spin-spin correlators also give insight into the magnetic ordering, the charge-charge correlators give an immediate quantitative signature of how far the system has entered the Spin Ice II phase. The charge-charge correlator is given by $C_{ij} = \langle Q_i Q_j \rangle / q^2$, which quantifies the degree of correlation between the signs of charges $Q_i$ and $Q_j$. Defining the location of the charges at representative vertices A, B, and C shown in Fig. 2(a), we determine the magnitude of $C_{ij}$ in the 8 × 8 µm² region for the first-nearest neighbors, $C_{AB}$, and second-nearest neighbors $C_{AC}$. In the ideal Spin Ice II phase, first nearest neighbors are perfectly anticorrelated ($C_{AB} = -1$) and second neighbors are perfectly correlated ($C_{AC} = +1$). For our laser annealed experimental systems, we find that $C_{AB} \sim -0.3$, indicating partial charge ordering, while $C_{AC} \sim +0.1$, consistent with a weak tendency towards the expected charge correlation of the Spin Ice II phase. These charge-charge correlators indicate the formation of vertex charge patterns consistent with partial charge crystallization [18], with the charge-ordered crystallites highlighted with gray shading in Fig. 2(b), and are comparable to those reported for DMI-assisted thermal annealing in disconnected kagome ASIs [19]. Crucially, our femtosecond laser-induced magnetic relaxation achieves a comparable degree of ordering without altering the intrinsic magnetic properties of the system with geometrical or material modifications, and at a much shorter timescale.

## IV. DETERMINISTIC GROUND STATE WRITING VIA SUBLATTICE-SELECTIVE LASER ANNEALING

### A. Sublattice selectivity with Cr capping

While we have shown that laser annealing can promote charge ordering to a certain extent in kagome ASI, accessing the long-range ordered ground state requires a more targeted strategy. We exploit the fact that, in the ground state, when the magnetic moments of the nanomagnets are projected onto a fixed in-plane axis, in this case the x-axis (see coordinate axes in Fig. 3), the lattice decomposes into two interleaved sublattices (distinguished by macrospins outlined and not outlined in black in Fig. 3) with macrospin components that point in opposite directions [see Fig. 3(d)]. Starting from a magnetic configuration with all macrospins pointing towards +x [see Fig. 3(a)], reversing the macrospins of only one of the two sublattices [macrospins shown in blue without an outline in Fig. 3(d)], yields the spin-crystal ground state in a single step. This insight motivates our experimental design: we selectively address one sublattice under a reversed sub-coercive field (i.e. the sublattice containing macrospins without a black outline) while the complementary sublattice (containing macrospins with a black outline) remains pinned.

To experimentally realize the site-selective switching using magnetically and geometrically equivalent nanomagnets, we introduced a differential optical absorption between the two sublattices via selective Cr capping. Specifically, one sublattice was fabricated with a standard Py(8 nm)/Al(3 nm) stack, while the other was modified by adding a 6 nm Cr capping layer above the Al layer, to give a Py(8 nm)/Al(3 nm)/Cr(6 nm) stack [Fig. 1(a)]. This configuration preserves the magnetic equivalence of the two sublattices, thus leaving the magnetic frustration unaltered. A potential concern when placing Cr in close proximity to Py is the possible presence of an exchange bias, which could affect the switching behavior. For this reason, the 3 nm-thick Al spacer was intentionally retained to prevent any direct magnetic coupling between Cr and Py. The presence of such a non-magnetic interlayer has been shown in previous studies to be sufficient to suppress exchange bias effects in similar systems [32]. To verify this, we measured the hysteresis loops of the Al/Cr-capped and Al-capped ASIs and found no evidence of shifted hysteresis loops or increased coercivity that is normally associated with exchange bias, thus confirming that exchange bias is not present within the limits set by the instrumental accuracy.

We now apply an experimental protocol to deterministically achieve the ground state, which involves varying the applied magnetic field and laser pulse energy over time as shown in Fig. 3(e). Before laser exposure, the entire system was initialized in a uniform magnetic state by applying a saturating magnetic field $H_{init}$ = 500 Oe along the *x* direction [Fig. 3(a)]. After removing the field, the magnetic moments of all nanomagnets remained aligned towards *x*, as illustrated in the Fig. 3(a). A reverse magnetic field $H_{rev}$ = -50 Oe, well below the ~200 Oe coercive field of the kagome ASI, was then applied and maintained during the laser annealing process [Fig. 3(b) and (c)]. The ASI was then exposed to the femtosecond pulsed laser beam with a fluence of 7.9 mJ/cm$^2$. We envisage that the reversal of the sublattice without the Cr layer proceeds as follows. At first, the magnetic moments of all nanomagnets still point towards *x* [Fig. 3(b)], since they are not sufficiently demagnetized to reverse under the applied sub-coercive field. As the laser pulse energy increased, only the Al-capped nanomagnets absorbed sufficient energy to reach thermal activation, enabling magnetization reversal of the macrospins without the black outline in Figs. 3(a-d) under the reversed sub-coercive field, while the magnetic moments in the Al/Cr-capped elements (with a black outline in Figs. 3(a-d)), maintained their original orientation. After the laser pulse was terminated and the applied magnetic field was removed [Fig. 3(d)], MFM imaging revealed that the scanned region (~20 × 20 μm$^2$), which is well within the laser spot size of ~85 μm (FWHM), relaxed into a nearly perfect spin-crystal ground state configuration. The reconstructed spin map, obtained from the analysis of the MFM magnetic contrast, is shown in Fig. 3(f), where ground state hexagonal plaquettes are highlighted in gray. White regions indicate defects where the hexagonal plaquettes are not in the ground-state configuration. These defects can be attributed to isolated fabrication errors such as shape irregularities or edge roughness that locally increase the energy barrier to switching. Since the defects normally involve the wrong orientation of a single spin, which is shared by two hexagonal plaquettes, the hexagonal plaquette defects shown in white usually come in pairs. A zoomed-in region in Fig. 3(f) displays the detailed magnetic configuration, which is overlaid in one corner with the corresponding MFM image, illustrating the correspondence between the measured contrast and the reconstructed spin map.

### B. Simulations of laser-induced heat transfer in uncapped and Cr-capped stacks

To further support the experimental findings, we performed simulations of laser-induced heat transfer in Py/Al and Py/Al/Cr stacks. The simulations were carried out using the *Heat Transfer in Solids* module in COMSOL Multiphysics® with a time-dependent study and fs time resolution. The incident laser beam was modeled as a Gaussian spot, with a full width half maximum of 85 μm as in the experiment, and with the individual pulses having a rectangular temporal profile of 100 fs duration, corresponding to the experimental fluence and repetition rate. The simulated geometry consisted of an 8 nm-thick Py film on a Si substrate, with and without a 6 nm Cr capping layer. Al layers were taken into account through effective interface boundary conditions but were not explicitly included as separate layers for computational efficiency. For this reason, they are denoted in parentheses for the Py/(Al) and Py/(Al)/Cr stacks when discussing the simulations. Since the laser penetration depth within the simulated time window is only ~0.6 μm, which is much smaller than the actual substrate thickness, the Si substrate could be truncated to 2 μm with its lower boundary maintained at ambient temperature. Heat exchange with the environment was included only through radiation from the topmost surface. The Py/Cr and Si/Py interfaces were modeled by introducing effective thermal boundary resistances in order to account for imperfect heat transmission across material discontinuities.

The absorption of the laser beam inside the exposed material was implemented following the Beer-Lambert Law [33,34], to give an exponential attenuation of the optical intensity with depth. From the simulations [Fig 4(a)], we see that, for the Py/(Al) stack, most of the optical energy is directly deposited in the magnetic (Py) layer, resulting in a sharp temperature rise across its thickness. In contrast, in the Py/(Al)/Cr stack, we see that a large fraction of the incident energy is absorbed within the Cr capping layer itself, both because of enhanced optical reflectivity at the Cr surface and because of its large intrinsic absorption coefficient at a laser wavelength of 515 nm. As a result, the heating of the Py underneath is significantly reduced and temporally delayed.

In more detail, the simulated depth-resolved temperature profiles through the stack, from the central pixel (lateral dimensions $5 \times 5$ nm$^2$) of the Gaussian laser spot, are shown in Fig. 4(a) for two different time delays. At $t_1 = 100$ fs, the temperature of the Cr capping layer in the Py/(Al)/Cr stack reaches a maximum of ~960 K (black solid line), while the temperature of the Py layer in the Py/(Al) stack attains a peak value of ~800 K (red solid line). The strong absorption of the laser light in Cr confines the energy near the surface, whereas the Py/(Al) stack distributes the energy more uniformly but less intensely. At $t_2 = 600$ fs, the temperature of the Py in the Py/(Al)/Cr stack reaches a maximum of ~675 K (black dashed curve), which is delayed with respect to the Py/(Al) stack because heat must first diffuse through the Cr capping layer. For completeness, the temperature profile of the Py/(Al) stack at $t_2 = 600$ fs is also shown (red dashed curve), although its maximum already occurred earlier.

Overall, the simulations demonstrate that the Cr capping layer does not simply act as a passive heat shield but modifies both the spatial and temporal energy absorption profile. We can deduce that, since the temperature of the Py in the Py/(Al)/Cr stack never exceeds the Curie temperature ($T_C \sim 770$ K [30]), the Py/Al/Cr nanomagnets retain a sufficient residual magnetization to prevent reversal under the applied sub-coercive magnetic field. In contrast, the simulations indicate that the Py/Al nanomagnets approach the Curie temperature and undergo substantial demagnetization, reducing the energy barrier to switching and enabling laser-assisted magnetization reversal. These simulation results therefore provide quantitative support for the mechanism underlying the deterministic writing of kagome ASI ground states, a method that is both robust and applicable to large areas, which is essential for programmable nanomagnetic architectures.

### C. Sublattice selectivity through modification of the nanomagnet thickness

Returning to the experiment, an alternative approach to Cr-induced optical attenuation for achieving sublattice selectivity is to modify the thickness of the magnetic material itself. In this implementation, one sublattice consists of standard Py/Al nanomagnets with a Py thickness of 8 nm, while the other sublattice is fabricated with an increased Py thickness of 14 nm. All other geometrical parameters remain identical.

COMSOL simulations of the temperature profiles [Fig. 4(b)], under the same laser fluence used in the Cr-capping approach, reveal that the internal temperature gradient is similar for both thicknesses. Nevertheless, due to the increased thickness, much of the thicker Py layer remains far below the Curie temperature throughout the pulse duration. As a result, a substantial portion of the volume retains a residual magnetization, which keeps the energy barrier to switching high enough so that the magnetization effectively remains frozen. In contrast, the thinner nanomagnets, with the temperature much closer to the Curie temperature, are almost entirely demagnetized, thus allowing laser-assisted reversal.

Experimentally, the final magnetic configuration obtained using this variable thickness approach again closely resembles the spin-crystal ground state, with long-range-ordered loops of macrospins visible across the scanned region [see Fig. 5].

Although this method does not preserve the magnetic equivalence of the two sublattices, and therefore lifts the intrinsic degeneracy of the kagome ASI, the use of thickness as a selection parameter offers the practical advantage of eliminating the need for additional capping materials. This strategy thus represents a viable and pragmatic alternative for implementations where robust and scalable optical site-selective switching is desired, while magnetic frustration is not a strict requirement. In this respect, this approach is interesting for the implementation of functional architectures, such as logic devices or reconfigurable magnetic systems.

## V. CONCLUSIONS

We have demonstrated a deterministic and rewritable approach to access the spin-crystal ground state of kagome ASI using ultrafast site-selective laser annealing. By tuning specific sublattices of the kagome ice geometry with targeted differences in optical absorption, we have enabled magnetization reversal of selected nanomagnets under sub-coercive magnetic fields. The modification of the local absorption is achieved through either Cr capping or changing the magnetic thickness, which is realized with standard lithography methods.

Our implementation based on capping specific subsets of nanomagnets with Cr, preserves the magnetic equivalence of the nanomagnets and maintains the intrinsic frustration. With this, it was possible to access the degenerate ground state manifold, as confirmed by MFM imaging and supported by heat transfer simulations confirming the sublattice selectivity. The alternative implementation, relying on thickness-based selectivity, offers a route to deterministic switching that only requires adjusting the nanomagnet thickness, so avoiding additional capping layers. Although this second method lifts the degeneracy of the low energy states because the two sublattices now have different magnetic properties, it is suitable for applications where maintaining the intrinsic frustration is not critical. Both approaches provided almost prefect long-range ordered magnetic states comparable to those obtained by modifying the Kagome ASI geometry.

In summary, we have established ultrafast laser annealing as a versatile tool for controlling frustrated nanomagnetic systems over large areas, without introducing changes to the lattice geometry. Indeed, while our current experiments are restricted to the size of the laser spot, the area of exposure can be easily increased by scanning the sample under the laser beam. Beyond ground-state writing, this method offers new opportunities for constructing input-selective nanomagnetic logic circuits and neuromorphic computing architectures, where different nanomagnets can be addressed at specific laser fluences. This approach also provides a route to design reprogrammable magnonic crystals with the ability to encode specific magnetic states, and to reset and re-established the magnetic configuration using the same laser excitation protocol, thus providing a means to precisely tune the spin-wave spectra during device operation at ultrafast timescales.

## ACKNOWLEDGMENTS

The authors acknowledge financial support from the Swiss National Science Foundation (Grant No. 200020_200332). We thank Gavin Macauley for developing the Python code used for the spin reconstruction analysis, Anja Weber and the PICO cleanroom staff at the Paul Scherrer Institute for technical support during the nanofabrication process, the members of the Mesoscopic Systems group for their assistance in the laboratory, and the Scanning Probe Microscopy Laboratory at the Paul Scherrer Institute for supporting the MFM measurements.

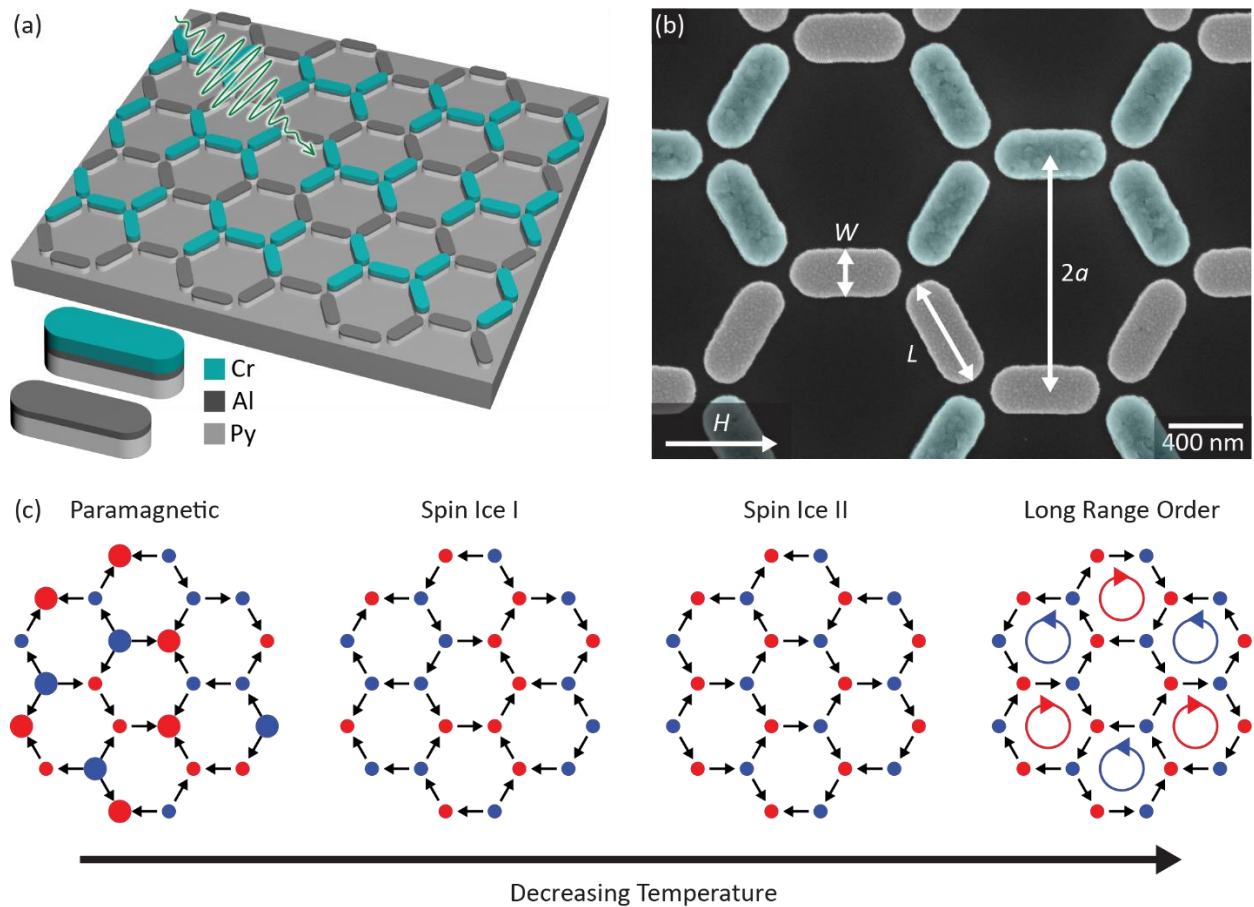

FIG. 1. Design of the kagome ASI, encoded with site-selectivity, and its magnetic phases. (a) Schematic of the kagome ASI consisting of Py nanomagnets fabricated on a Si substrate and capped either with Al or with Al/Cr. The Cr capping layer (shown in turquoise) has a higher optical absorption than the Al capping layer (shown in dark gray), enabling sublattice-selective laser activation while maintaining magnetic equivalence between the sublattices. (b) Scanning electron microscopy image of the kagome ASI, consisting of permalloy nanomagnets with length $L$ = 300 nm, width $W$ = 100 nm, and lattice constant $a$ = 320 nm. The center-to-center distance between nanomagnets across a hexagonal plaquette is $2a$. Nanomagnets capped with an Al/Cr bilayer are highlighted in turquoise. The orientation of the applied magnetic field $H$ is also indicated. (c) Schematics of theoretically predicted magnetic phases of kagome ASI obtained on cooling. Small red (blue) dots correspond to ice-rule-obeying 2-in/1-out (1-in/2-out) vertex configurations, while large red (blue) dots indicate high-energy 3-in (3-out) defects. In the high-temperature paramagnetic regime, macrospins fluctuate without correlations. As the temperature is reduced, the Spin Ice I phase is reached where the ice rule is obeyed at every vertex. This is followed by the Spin Ice II phase, which exhibits long-range ordering of vertex charges, where neighboring charges alternate to minimize the Coulomb-like interaction energy. At the lowest temperatures, dipolar interactions stabilize the long-range ordered ground state, consisting of alternating clockwise (blue arrows) and counterclockwise (red arrows) loops of macrospins.

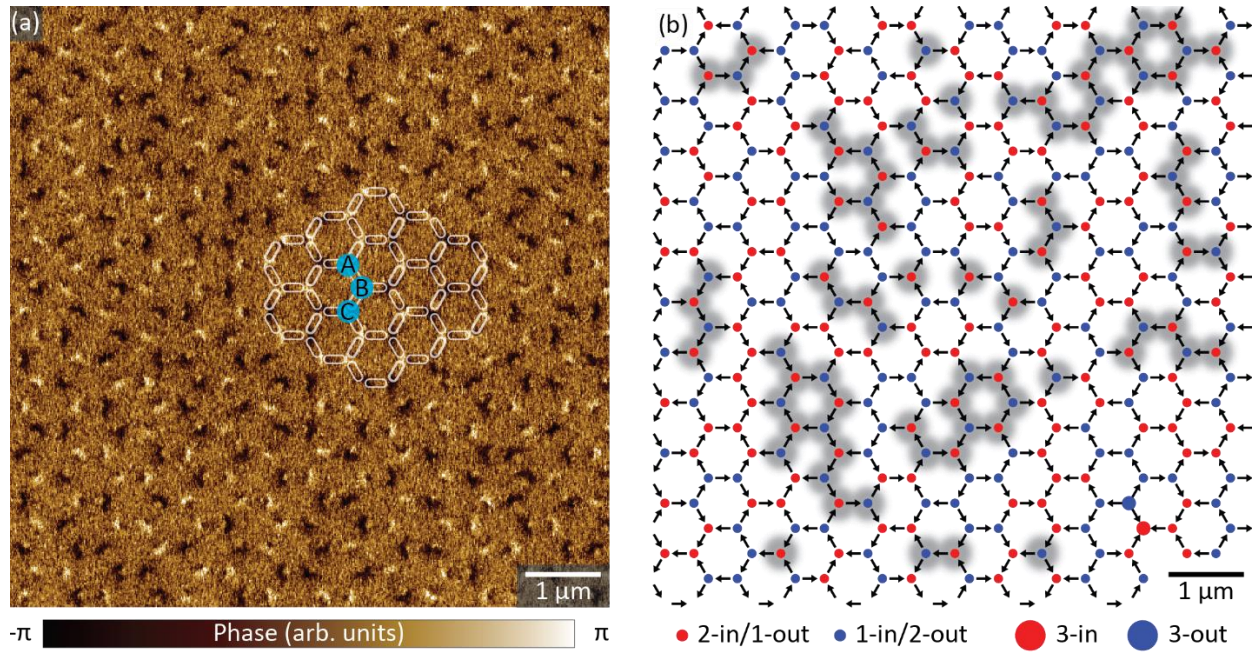

FIG. 2. Magnetic configuration of kagome ASI after ultrafast laser annealing revealed by MFM imaging and reconstructed vertex charge map. (a) MFM image of a kagome ASI following ultrafast laser annealing. Bright (dark) contrast corresponds to the magnetic stray field associated with the north (south) magnetic poles at the ends of the nanomagnets. Three representative vertices are highlighted (A, B, C), illustrating how nearest-neighbor (A-B) and next-nearest-neighbor (A-C) charge correlators are defined. (b) Reconstructed macrospin and vertex charge map obtained from the MFM image in (a), where small red (blue) dots correspond to the +q(-q) charges associated with 2-in/1-out (1-in/2-out) ice-rule-obeying vertices, and large red (blue) dots indicate the +3q(-3q) charges associated with high-energy 3-in (3-out) defects. Unit vertex charges surrounded by three unit charges of opposite sign are highlighted with gray shading revealing partial charge crystallization consistent with the onset of the Spin Ice II phase.

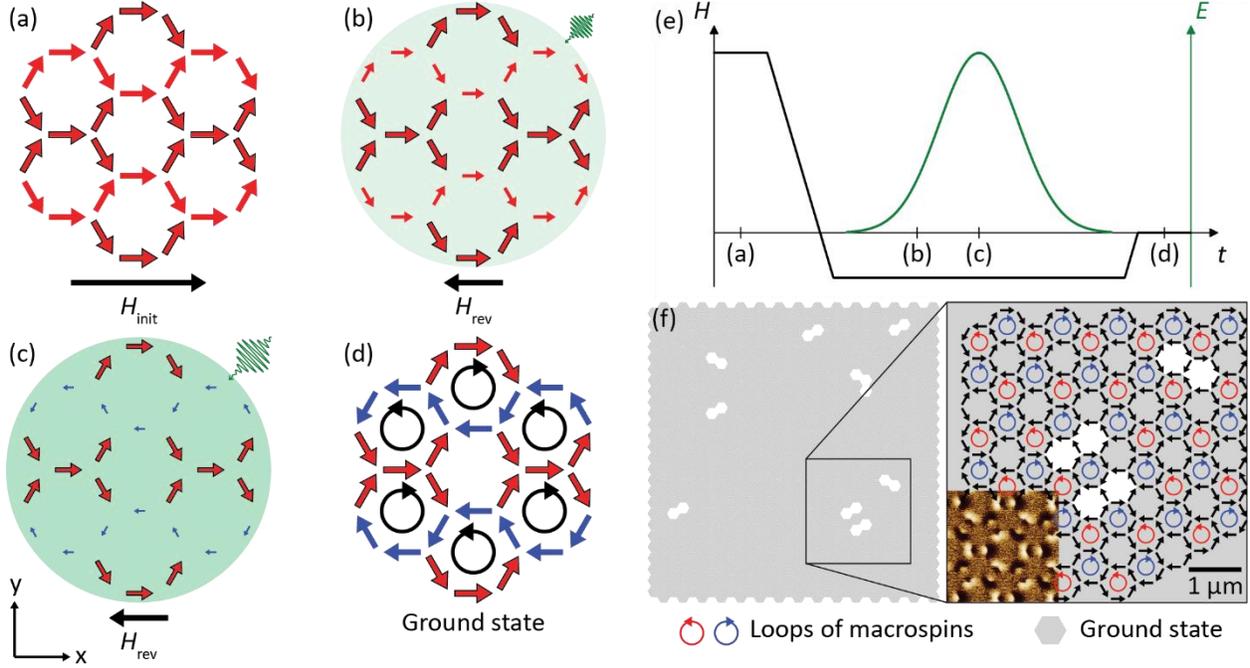

FIG. 3. Schematic representation and experimental realization of deterministic ground-state writing in kagome artificial spin ice via sublattice-selective ultrafast laser annealing. (a) The system is first initialized by applying a saturating magnetic field $H_{init}$ = 500 Oe along the $x$ direction so that, on removal of the field, all macrospins to point towards the right. Nanomagnets outlined in black correspond to the Al/Cr-capped sublattice, while Al-capped elements form the complementary sublattice. (b) A reverse magnetic field of $H_{rev}$ = 50 Oe is then applied, which is well below the ~200 Oe coercive field of the nanomagnets, and the laser exposure (shown with green shading) begins to demagnetize the nanomagnets. (c) At a later time during laser exposure, the Al-capped nanomagnets absorb sufficient energy to become sufficiently demagnetized to switch their residual magnetization under the reverse field, while the magnetization in the Al/Cr-capped elements remains large enough to prevent reversal. (d) After the laser pulse terminates, the system's magnetization recovers into the fully-ordered spin-crystal ground state, maximizing the number of alternating clockwise and counterclockwise loops of macrospins. (e) Schematic representation of the experimental protocol showing how the applied magnetic field $H$ and the laser pulse energy $E$ change over time $t$. The labels (a) to (d) indicate the stages of the corresponding figure panels. (f) Spin configuration obtained with MFM over a 20 × 20 µm² region, which is well within the laser spot that has a size of ~85 µm. This confirms the formation of long-range order (hexagonal plaquettes indicated in gray) and only a few defects i.e. deviations from the ground state (hexagonal plaquettes shown in white), which are likely to be due to local fabrication irregularities. A fully-ordered region of the original MFM image has been included to illustrate the correspondence between the measured contrast and the reconstructed configuration.

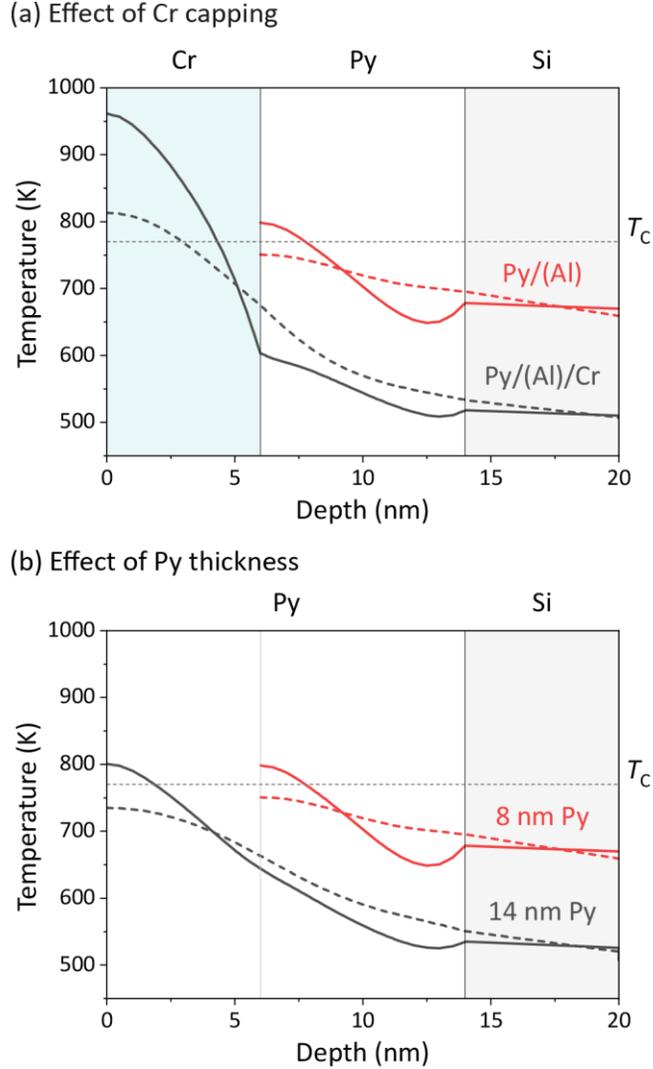

FIG. 4. Simulated temperature - depth profiles of Py/Al/Cr and Py/Al film stacks under femtosecond laser excitation. (a) Depth-resolved temperature distribution for Py nanomagnets with (black curves) and without (red curves) a Cr capping layer at two different times after the laser pulse: $t_1$ = 100 fs (solid lines) and $t_2$ = 600 fs (dashed lines). In the Cr-capped stack, most of the incident energy is absorbed in the Cr layer, where peak temperature exceeds 900 K at $t_1$, while the underlying Py reaches a reduced peak temperature of about 675 K at the later time $t_2$, and remains well below the Curie temperature $T_C \sim 770$ K. In contrast, in the Py/(Al) stack, the energy is directly deposited in the magnetic layer, which rapidly approaches $T_C$, enabling substantial demagnetization. Brackets for the Al layers indicate that they are not modeled as explicit layers but are instead included via effective interface boundary conditions. (b) Depth-resolved temperature distribution for different thicknesses of Py (8 nm and 14 nm) at two different times after the laser pulse: $t_1$ = 100 fs (solid lines) and $t_2$ = 600 fs (dashed lines). In both cases, the temperature profiles decrease rapidly with depth. The thinner magnetic layer reaches temperatures close to $T_C$ across the thickness whereas, in the thicker layer, only the upper region is significantly heated, leaving a large fraction of the magnetic layer well below $T_C$. This difference in thermal response supports the experimental results where the thinner nanomagnets undergo switching, while the thicker nanomagnets remain in their initial state during laser excitation.

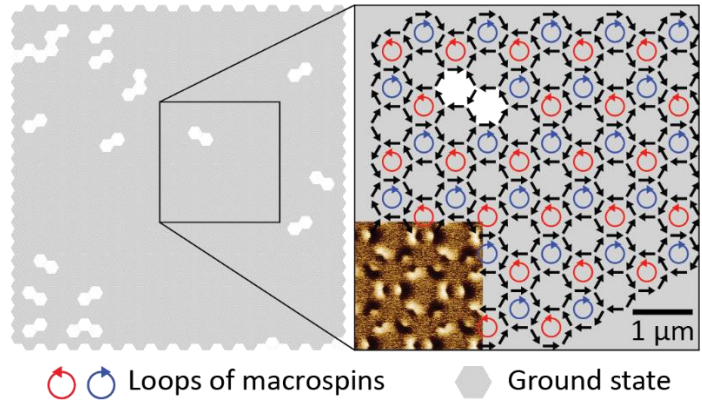

FIG. 5. Spin configuration of kagome ASI obtained via sublattice-selective ultrafast laser annealing based on different Py thicknesses in the two sublattices (8 nm and 14 nm). The configuration is reconstructed from the original image obtained with MFM over a 20 × 20 µm$^2$ region, which is well within the laser spot, which has a size of ~85 µm. Similar to the Al/Cr capping approach shown in Fig. 3(f), the reconstruction confirms the presence of long-range order (shown in gray), characteristic of the spin-crystal ground state, with only a few defects (shown in white), most likely due to local fabrication irregularities. A fully ordered region of the original MFM image is included to illustrate the correspondence between the measured contrast and the reconstructed configuration.